# OVERVIEW OF LLRF SYSTEM FOR IBNCT ACCELERATOR

Z. Fang, K. Futatsukawa, Y. Fukui, T. Obina, Y. Honda, F. Qiu, T. Sugimura, S. Michizono, S. Anami, F. Naito, H. Kobayashi, T. Kurihara, M. Sato, T. Miyajima, KEK, Japan
T. Ohba, N. Nagura, Nippon Advanced Technology CO., LTD., Japan

*Abstract*

At the Ibaraki Neutron Medical Research Center, an accelerator-based neutron source for iBNCT (Ibaraki - Boron Neutron Capture Therapy) is being developed using an 8-MeV proton linac and a beryllium-based neutron production target. The proton linac consists of an RFQ and a DTL, which is almost the same as the front part of J-PARC linac. However, here only one high-power klystron is used as the RF source to drive the two cavities, which have quite different Q-values and responses. From June 2016, a cPCI based digital feedback system was applied to the iBNCT accelerator. It serves not only as a controller for the feedback of acceleration fields, but also as a smart operator for the auto-tuning of the two cavities in the meantime, especially during the RF startup process to the full power. The details will be described in this report.

## INTRODUCTION

The accelerator-based iBNCT (Ibaraki - Boron Neutron Capture Therapy) is being developed under collaboration of KEK and Tsukuba University. Now the construction of the iBNCT accelerator has been finished, and the beam commissioning has started. Figure 1 shows the outline of the iBNCT accelerator. Its layout is shown in Fig. 2.

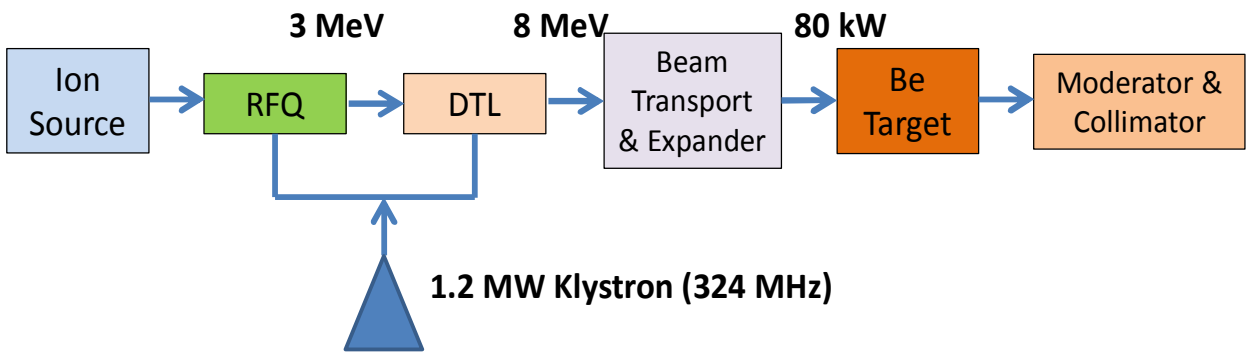

Figure 1: Outline of the iBNCT accelerator.

This is an 8-MeV proton linac, which is almost the same as the front part of J-PARC linac. At iBNCT, an RFQ and a DTL are used for the proton beam acceleration to 3 MeV and 8 MeV respectively. Figure 3 shows a picture of the RFQ and DTL cavities. The iBNCT accelerator is operated in pulse mode with a maximum repetition rate of 200 Hz. The maximum peak current of proton beam is designed as 50 mA, and the pulse width is 1 ms. So, the maximum beam power is 80 kW.

In order to reduce the cost of construction, only one high-power klystron is used as the RF source to drive the two cavities. A picture of the klystron and power divider is shown in Fig. 4. While as known, the RFQ and DTL cavities have quite different Q-values and responses. The loaded Q-values are about 3,500 and 16,000 for RFQ and DTL respectively. Therefore, how to operate this RF system with a good stability and a fast response is a very important issue for the low-level RF control system.

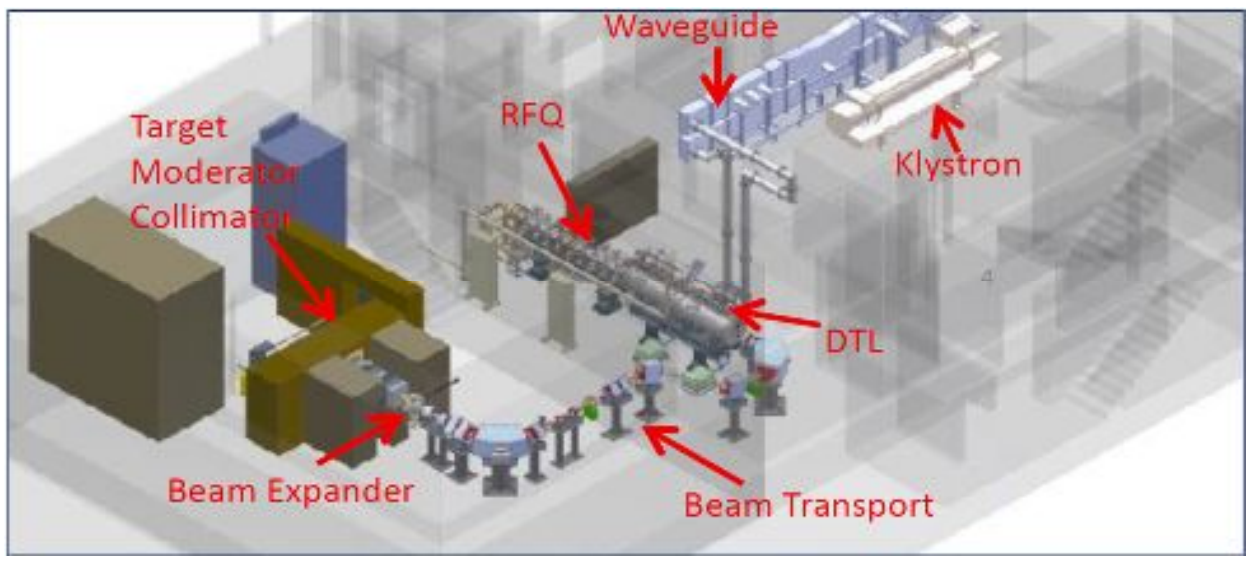

Figure 2: Layout of the iBNCT accelerator.

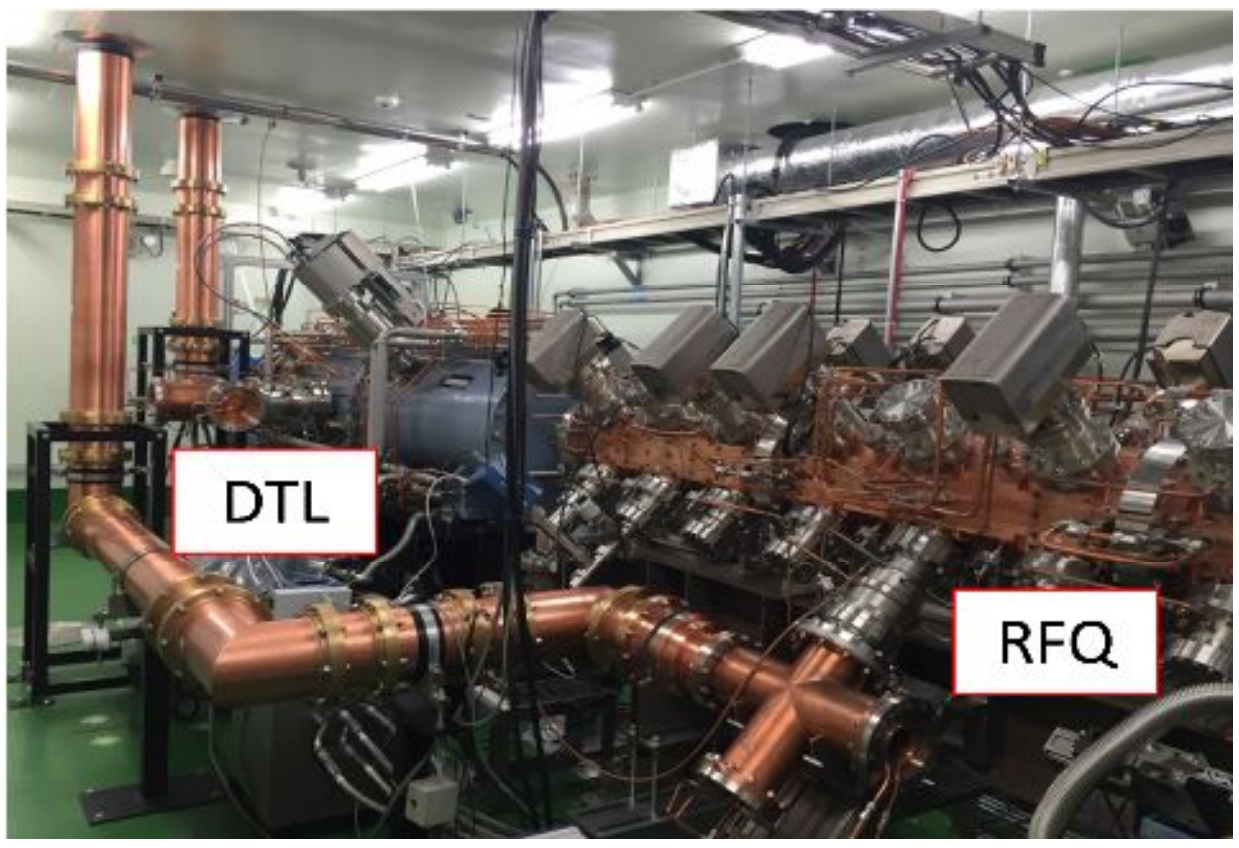

Figure 3: RFQ and DTL cavities at iBNCT.

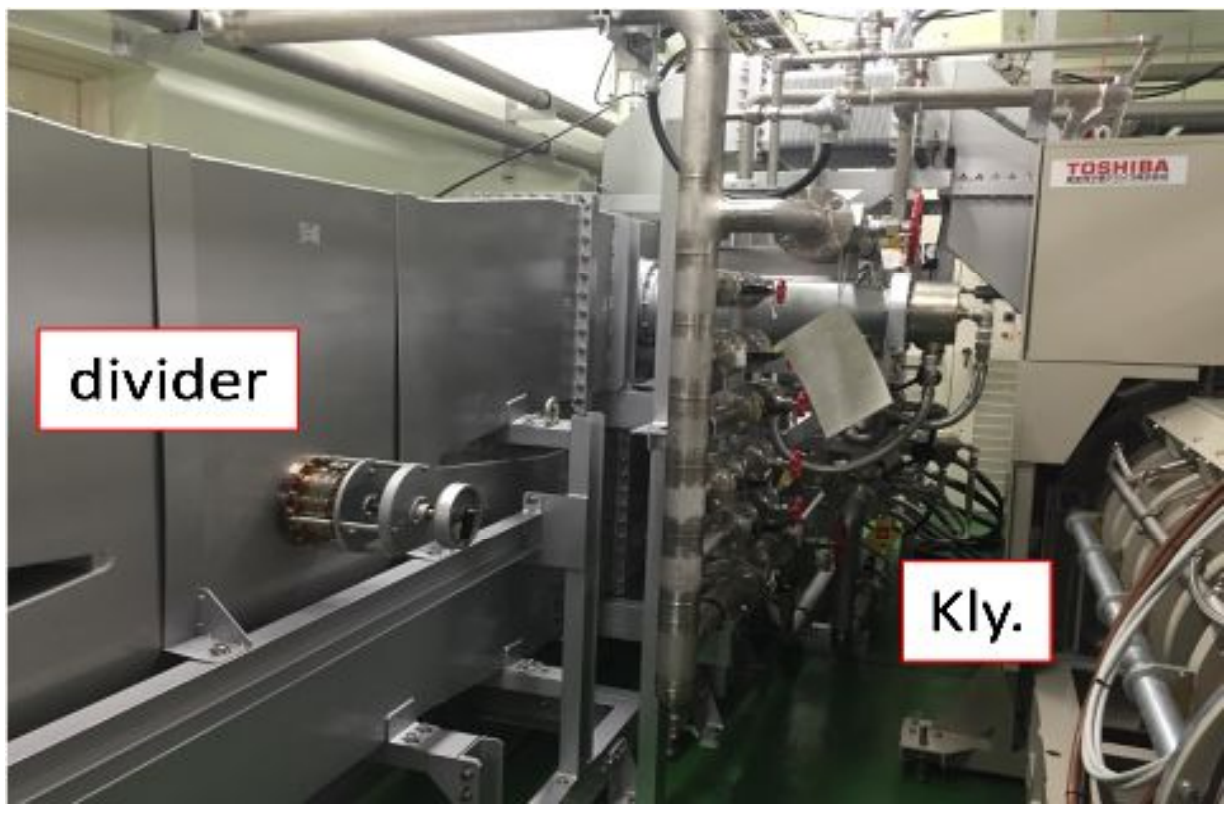

Figure 4: Klystron and power divider at iBNCT.

## LLRF SYSTEM

From June 2016, a cPCI based digital feedback system was applied to the iBNCT accelerator, which is also most the same as the LLRF system for the J-PARC linac [1-2]. The block diagram of RF system for iBNCT is shown in Fig. 5. A picture of the LLRF system is shown in Fig. 6. The RF driving power will be controlled in the digital feedback system, and delivered to the klystron via a 40-W pre-amplifier. The RF power of wall loss for the RFQ and DTL is 340 kW and 320kW respectively. The beam power gained from the RFQ and DTL cavities is about 150 kW and 250 kW respectively, at beam current of 50 mA. Totally the driving power from the klystron is about 1.06 MW. The cavity temperatures are controlled using a cooling water system. And for the DTL cavity, we have a mechanical tuner to control the DTL cavity resonance frequency precisely. The LLRF system serves not only as a controller for the feedback of acceleration fields, but also as a smart operator for the auto-tuning of the two cavities in the meantime, especially during the RF startup process.

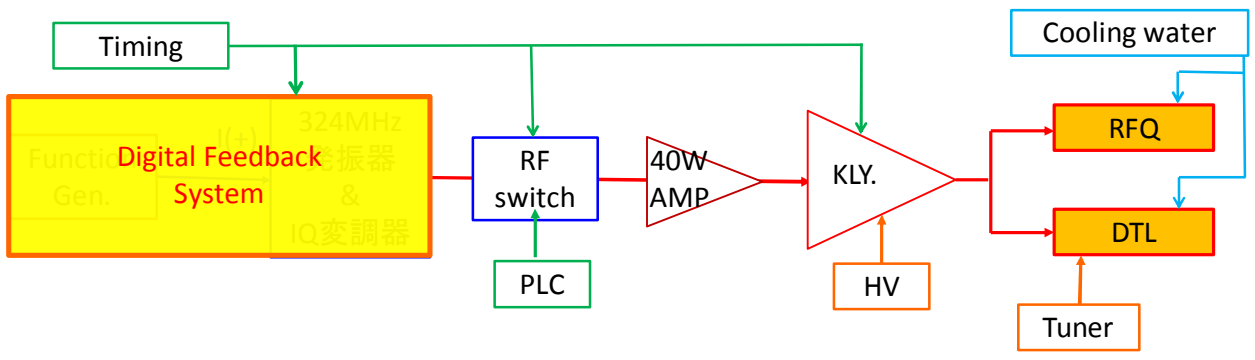

Figure 5: Block diagram of RF system for iBNCT.

## DTL TUNER CONTROLLER

In order to operate the accelerator with a good cavity tuning for both the RFQ and DTL cavities, a special DTL tuner controller has been developed, as shown in Fig. 7. There are three control loops.

The first control loop is used for tuning the DTL cavity to have a resonance frequency as the operation frequency 324MHz. For the J-PARC linac, we have only this first control loop to maintain the cavity resonance at operation frequency.

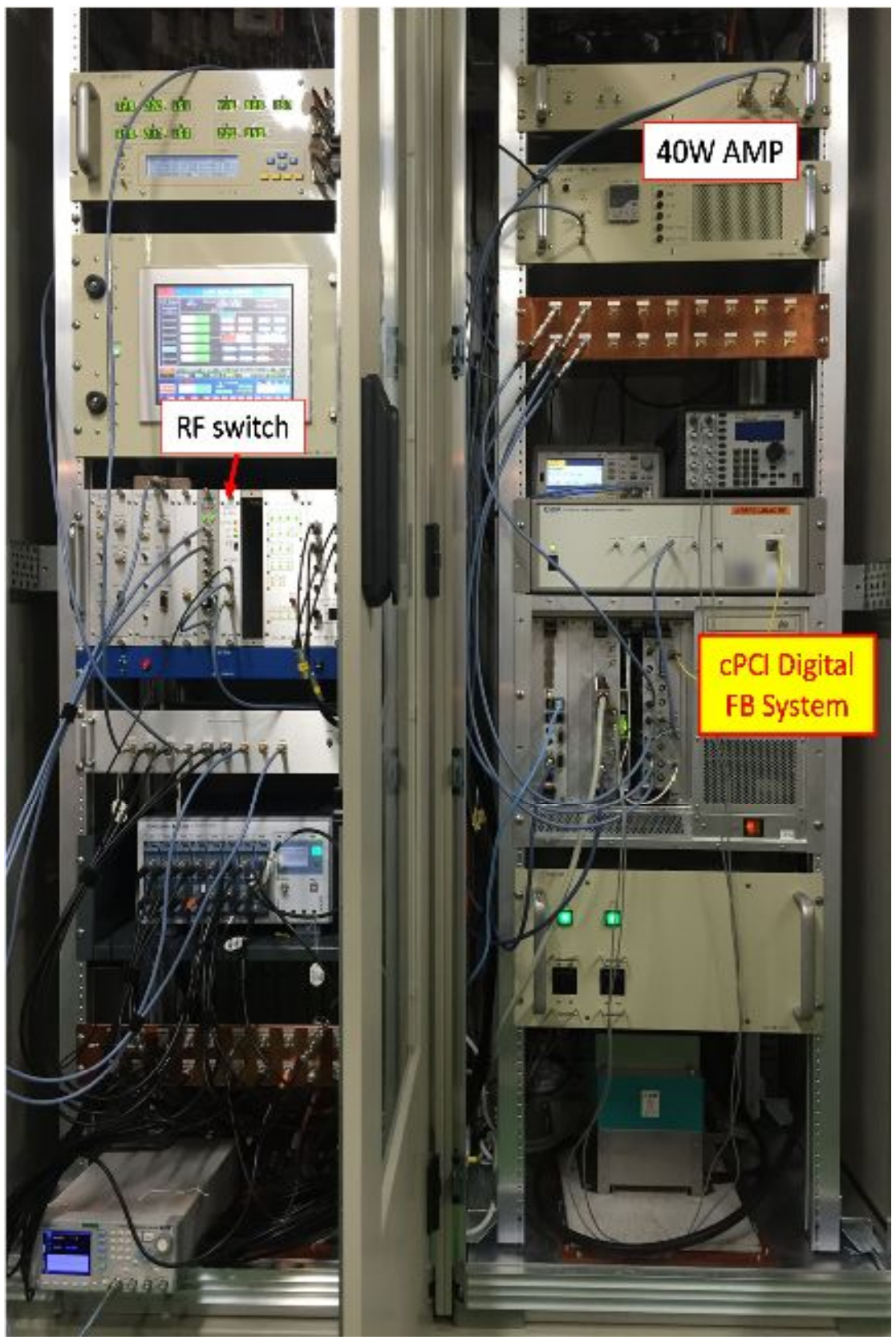

Figure 6: LLRF system of iBNCT.

At iBNCT, the second and third control loops will also be used. At the auto-startup process, the input RF frequency will be tuned to match the DTL cavity, so that we could warm up the cavities as soon as possible and restart the machine operation with beam acceleration quickly. During this warming up process, the second control loop is used for tuning the DTL cavity to have the same detuned frequency as RFQ. In this way, the input RF driving frequency, the resonance frequency of RFQ, and the resonance frequency of DTL, will be the same value. And the reflection from the two cavities will be minimized.

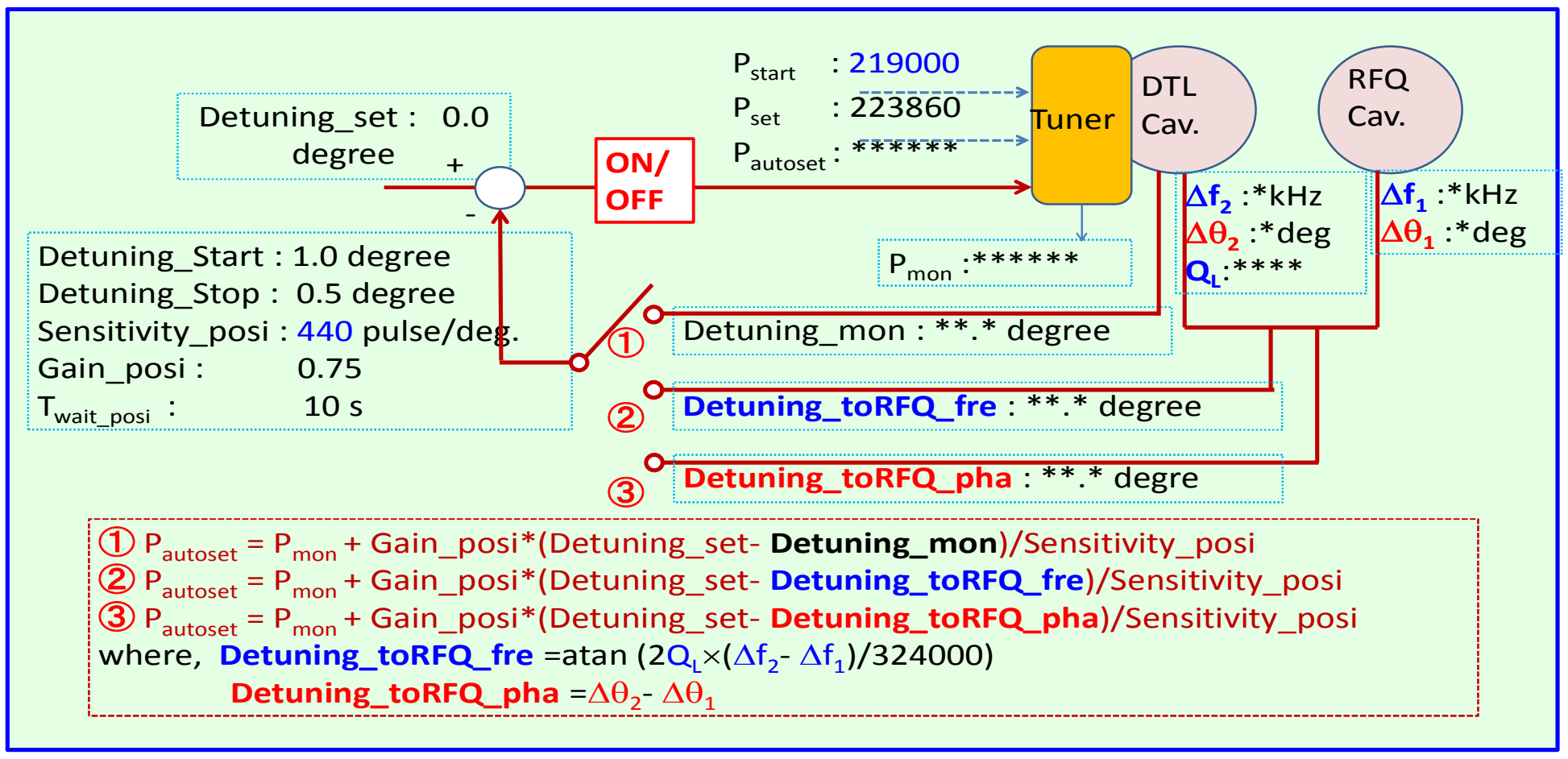

Figure 7: DTL tuner controller.

After the cavities warming up, the full required power will be fed to the cavities and a feedback control of acceleration fields will be turned on. However, we have only one driving source for the two cavities. After feedback on, the third control loop is used for tuning the DTL cavity to have the same detuned phase of RFQ. In this way, the two cavities will have the same impedance features including the amplitude and phase. The RFQ and DTL cavities will perform like one cavity. The RF fields (amplitudes and phases) will be kept stable for both the two cavities even the feedback control is turned on for one of the cavities.

## RF AUTO-STARTUP PROCESS

For the iBNCT accelerator operation, a special RF auto-startup process has been carried out. Figure 8 shows an example of experiment results for the RF auto-startup process. In this experiment, the accelerator is operated at a repetition rate of 75 Hz with the RF pulse width of 1 ms. At first, a little power is fed to the cavities with the input RF frequency tuned to the DTL cavity, and the DTL tuner is controlled to tuning the DTL cavity to have the same detuned frequency as RFQ, using the second control loop. Then the RF power will be increased quickly to the required power to warm up the RF cavities, while keeping the resonance frequencies of RFQ and DTL have the same value. When the RFQ cavity is warmed up to get resonance frequency close to the operation frequency, the DTL tuner control will be switched from the first control loop to the second control loop, so that to make the DTL cavity have a resonance at the operation frequency. When both the two cavities have a resonance at the operation frequency, then we could exchange the input RF frequency to the operation frequency and turn the feedback on and the quick recovery function on. The total time for this whole auto-startup process is about 6 minutes.

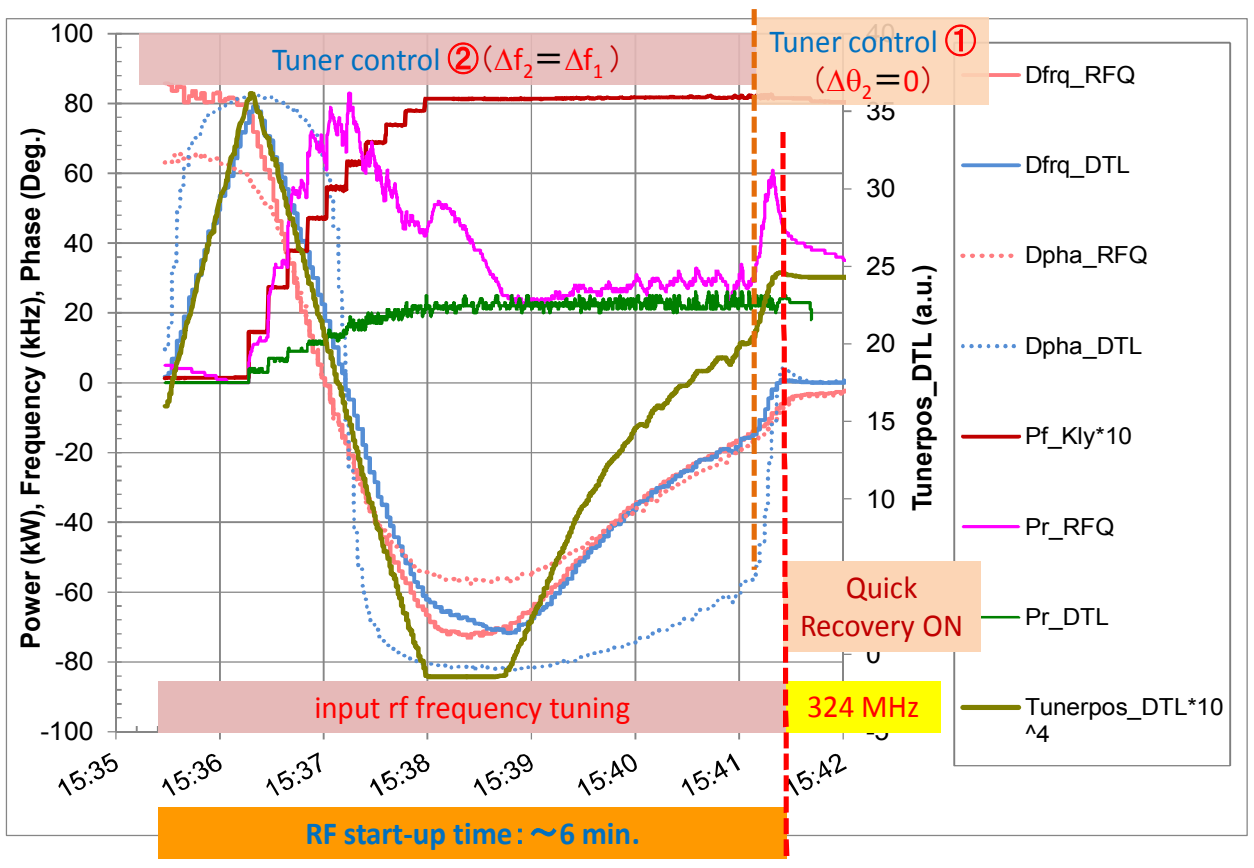

Figure 8: Auto-startup process.

## RF STABILITIES

Now the beam commissioning of iBNCT accelerator is going well. A proton beam with a peak current of 24 mA and pulse width of 850 μs has been successfully accelerated with very good stabilities. Figure 9 shows the RF waveforms of amplitude and phase at the flat top. The stabilities of amplitude for the beam acceleration at the flat top are about ±0.4% and ±0.3% for the RFQ and DTL respectively. And the stabilities of phase are about ±0.2° for both the RFQ and DTL cavities.

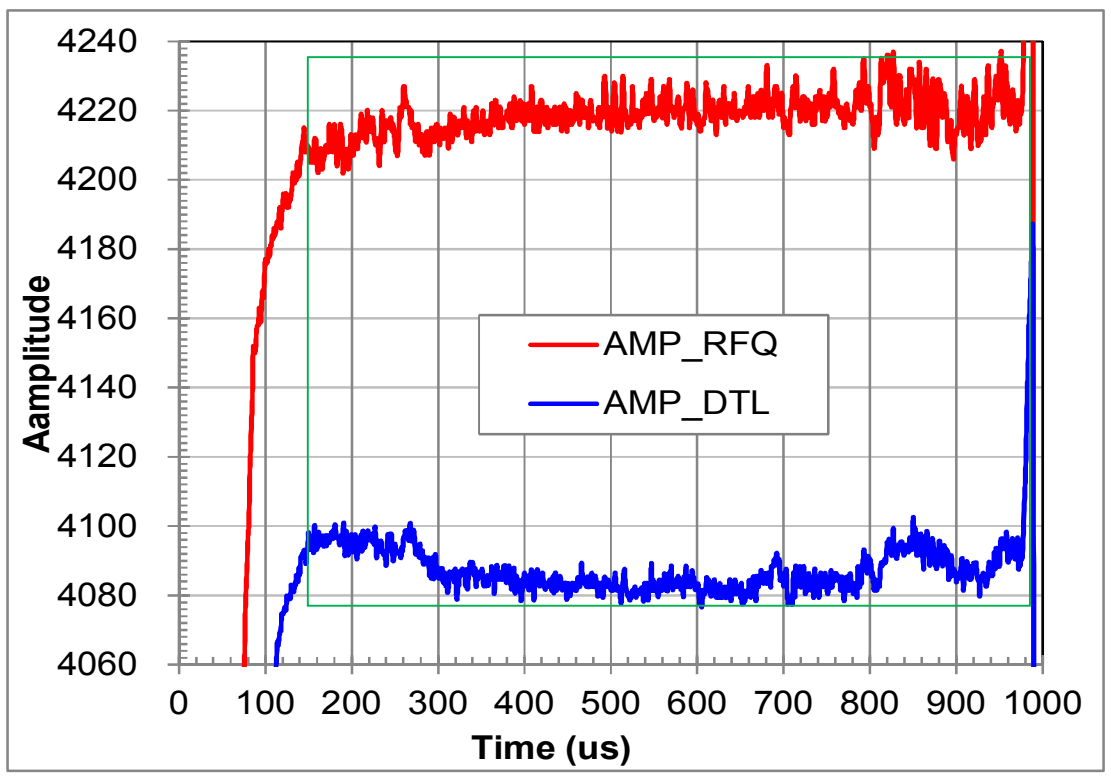

a) waveforms of amplitude at the flat top.

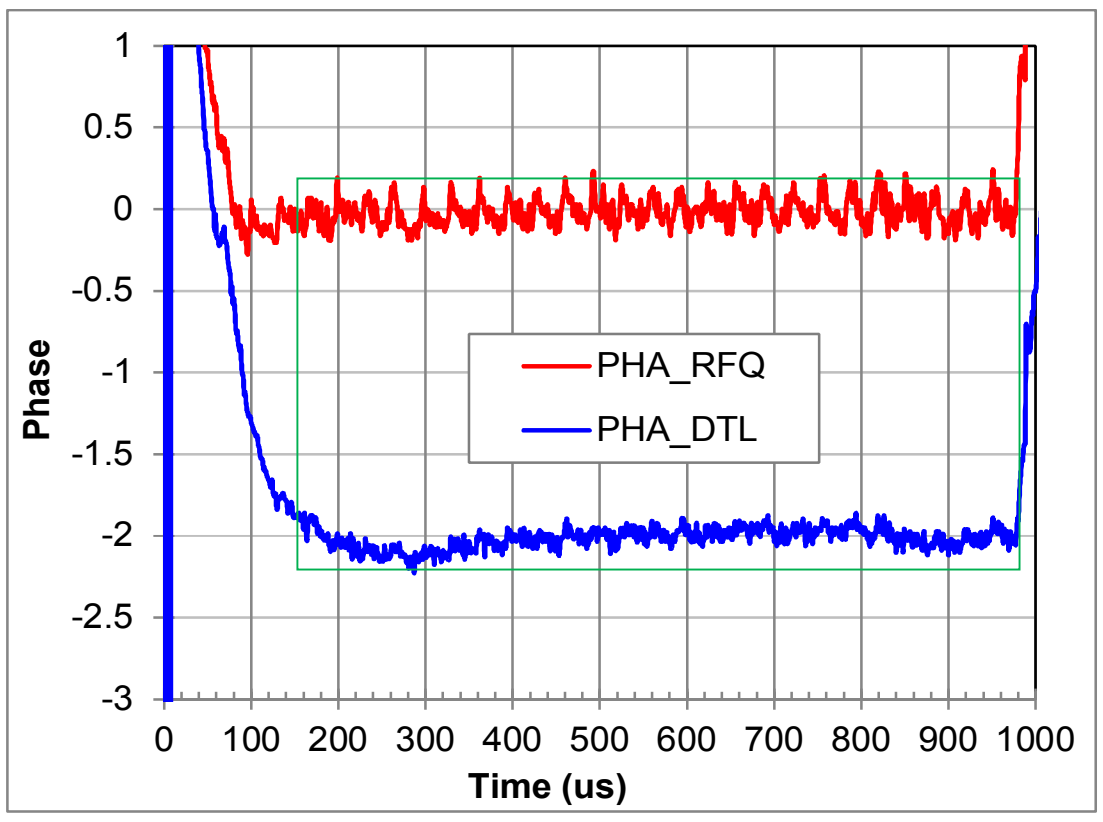

b) waveforms of phase at the flat top.

Figure 9: RF waveforms at flat top.

## SUMMARY

A digital feedback LLRF control system has been successfully applied at iBNCT with all the control functions same as J-PARC linac. An ultimate auto-tuning system including the DTL tuner controller for iBNCT has been developed. An auto-startup process for iBNCT has been worked out minimizing the reflection power from the RF cavities and the time of RF restart operation. Good performances of beam operation have been achieved with very good RF field stabilities.

## REFERENCES

[1] S. Anami et al., "Control of the Low Level RF System for the J-PARC Linac", LINAC 2004, 739–741, Germany.

[2] Z. Fang al., "Auto-tuning Systems for J-PARC Linac RF Cavities", Nuclear Instruments and Methods in Physics Researcher A (NIM. A), Vol. 767, 135-145, Dec. 2014.